\documentclass{article}

\usepackage{graphicx}
\usepackage{epsfig}
\usepackage{color}
\usepackage{url}

\begin{document}

\title{Opinions and dark energy: A mechanism of long-range repulsion in a toy opinion particle model}

\author{Andr\'e C. R. Martins \\ 
	GRIFE -- EACH -- Universidade de S\~ao Paulo,\\
	Rua Arlindo B\'etio, 1000, 03828--000,  S\~ao Paulo, Brazil
}

\maketitle              

\begin{abstract}
The concept of opinion particles can be introduced by studying time-continuous versions of Bayesian-inspired opinion dynamics methods. Here, we use opinion particles to further explore how information and Bayesian methods can contribute new ideas to physical problems. Continuous-time limits are shown to correspond to the case with little trust between the original agents, and a synchronous version of the model is studied. The interaction between two particles is defined as attractive. However, space and distances are not well represented in the model. We define a local measure of distances based on the parameters. That allows distant particles to be repelled in the natural coordinates of any central particle, similar to the observed cosmological effects identified as dark energy.

Keywords: Opinion dynamics, Opinion particles, Dark energy, Bayesian methods, CODA
\end{abstract}

\section{Introduction}

Opinion dynamics models~\cite{castellanoetal07,latane81a,galametal82,galammoscovici91,sznajd00,deffuantetal00,martins08a,martins12b,galam12a} have been traditionally used to explore social problems, such as convergence to consensus, the appearance of polarization and echo-chambers, as well as the emergence of extremist behavior. Using ideas from physics to create or improve opinion models is a well-established tradition \cite{galametal82,galam12a,schweitzerholyst00,Wang2017,Wang2021}. On the other hand, contributions from opinion models to general discussions are rare, but they do exist \cite{martins15b,galammartins15a}.

In this paper, we will further explore the idea of opinion particles \cite{martins15b} that arose as an application of Bayesian-inspired opinion dynamics models. Those opinion dynamic models were initially inspired by the "Continuous Opinions and Discrete Actions" (CODA) model and the framework that was proposed based on those ideas \cite{martins08a,martins12b,Martins2021}. Indeed, using concepts of information theory to understand physical ideas better is a well-established and promising approach \cite{jaynes03,caticha04a,catichapreuss04a,catichagiffin07a,goyal12a,Leandre2018}. This work aims to explore those possibilities further.

Opinion particle models based on CODA and on an altered version of CODA with continuous communication were shown before as an alternative way to define inertia and also as a version of the harmonic oscillator \cite{martins15b}.  Here, we will explore a continuous time extension of the Bayesian model for updating opinions about a continuous variable \cite{martins08c,Maciel2020}. That model was qualitatively similar to the well-known Bounded Confidence models of opinion dynamics \cite{deffuantetal00,hegselmannkrause02}, due to the introduction of the possibility of mistrusting the information from other agents. 

In the model, two parameters are updated, the mean position and the uncertainty of each agent/particle. However, the scale for those values is arbitrary. Here, we investigate what happens when the particle uses its uncertainty parameter, measured by the standard deviation of its opinion distribution, as a ruler to create a local, natural system of coordinates. In those coordinates, close particles still get attracted, but we observe the appearance of a repulsive force, akin to dark energy effects in cosmology \cite{Peebles2003}. While there are repulsion effects in opinion dynamics \cite{VazMartins2010}, including contrarians \cite{galam04,martinskuba09a,Khalil2019}, the repulsion in this toy model happens as a consequence of how the measurement is defined and it is not introduced ad-hoc.

\section{The model}

In the original model \cite{martins08c,Maciel2020}, the agents wanted to reach a consensus about a quantity $\theta$ and they assumed other agents might have relevant information about the value of $\theta$, where, in principle, $0 \leq \theta \leq 1$. That is, each agent $i$ thought there was a chance $p$ that agent $j$ estimate, $x_j$ would be around the correct value for $\theta$. And, since $j$ might not have a clue, there was also a $1-p$ probability that $j$ estimate was just random. To represent that, a uniform distribution term was introduced in the probability estimate for how likely $j$ would pick value $x_j$, assuming the correct value was $\theta$. In Bayesian terms, the likelihood of $x_j$ given $\theta$ was

\begin{equation}\label{eq:likelihooddecept}
	f(x_j|\theta) = p N(\theta,\sigma_{j}^{2}) + (1-p) U(0,1),
\end{equation}
where $N(\theta,\sigma_{j}^{2})$ represents the Normal distribution centered around $\theta$, with $sigma_{j}$ measuring how close or distant $j$ was expected to be from the actual value.

An update rule can be obtained using this likelihood as long as we have a prior opinion. The simplest choice, in this case, is to use a normal $N(\theta,\sigma_{i}^{2})$ for the opinion of agent $i$, that is

\[
f(\theta)=\frac{1}{\sqrt{2\pi}\sigma_i} e^{-\frac{(\theta-x_i)^2}{2\sigma_i^2}}.
\]

Bayes theorem is nothing more than multiplying that prior by the likelihood and renormalizing the result so that probabilitities will integrate to one over the whole space. From that, when agent $i$ observes agent $j$ estimate, $x_j$, we get the posterior
\begin{equation}\label{eq:posterior}
	f(\theta|x_j)\propto p e^{-\frac{1}{2\sigma_{i}^{2}}[(\theta-x_i)^2+(x_j-\theta)^2]}+(1-p) e^{-\frac{(x_i-x_j)^2}{2\sigma_{i}^{2}}}.
\end{equation}

To simplify the problem, we assume agents only update their average estimate $x_i$ and their uncertainty $\sigma_i$. From Equation~\ref{eq:posterior}, we get the update rules 

\begin{equation}\label{eq:averagedecept}
	x_i(t+1)=p^*\frac{x_i(t)+x_j(t)}{2}+(1-p^*)x_i(t)
\end{equation}
where
\begin{equation}\label{eq:posteriorp}
	p^* = \frac{p\frac{1}{2\sqrt{\pi}\sigma_i} e^{-\frac{(x_i(t)-x_j(t))^2}{4\sigma_{i}^{2}}} }{p\frac{1}{2\sqrt{\pi}\sigma_i} e^{-\frac{(x_i(t)-x_j(t))^2}{4\sigma_{i}^{2}}} +(1-p)},
\end{equation}

and

\begin{equation}\label{eq:varupdate}
	\sigma_{i}^{2}(t+1)=\sigma_{i}^{2}(t)\left( 1-\frac{p^*}{2} \right) + p^*(1-p^*) \left(\frac{x_i(t)-x_j(t)}{2}\right)^2.
\end{equation}

It is worth noticing that Equation~\ref{eq:posteriorp} was printed with an error in numerical constants in the original papers \cite{martins08c,Maciel2020}.

\subsection{Opinion particles version of the update rule}

To obtain a model we can describe as the movement of an opinion particle from Equations~\ref{eq:averagedecept} and \ref{eq:varupdate}, we need to take the limit of minimal updates, corresponding to the limit of continuous time. That was accomplished before by assuming each discrete update corresponds to a finite and non-zero time step and investigating how smaller steps could lead to the continuous limit \cite{martins15b}. Here, however, we can make the changes in each time step arbitrarily smaller simply by decreasing the value of the parameter $p$. When $p$ tends to zero, from Equation~\ref{eq:posteriorp}, we can see that $p^*$ also tends to zero. The effect of that is that the changes in the mean and standard deviation of the posterior become negligible, as they should be, for well-behaved functions, in the limit where the time step tends to zero. We can use a non-zero but small value of $p$ to generate an approximation to the actual continuous limit for simulation purposes.

Another distinction from the opinion update rule is necessary here. When the model was used for social simulation, using an asynchronous update rule made sense. Discussions happen one at a time. But that is not necessarily true for particles. It makes sense to have them interact simultaneously. To introduce that, we can, for each opinion particle $i$, use Equation \ref{eq:averagedecept} to calculate the change $c_{ij} = x_i(t+1)- x_i(t)$ each other particle $j \neq i$ causes in $i$ and add them all to get the effective new position of the average 
\begin{equation}\label{eq:meancont}
	x_{i}(t+\Delta t)= x_i(t) + \sum_j c_{ij},
\end{equation}
where the change in time is represented by $\Delta t$ to make it clear that this is a continuous version and also to signal we have not defined a proper time scale.

For the standard deviations $\sigma_i(t)$, adding all changes from Equation \ref{eq:varupdate} could easily cause a non-realistic negative value. Instead, it makes more sense to use the update equation to estimate ratio changes $s_{ij} = \frac{\sigma_i(t+1)}{\sigma_i(t)}$, so that the total synchronous change can be given by
\begin{equation}\label{eq:uncertcont}
	\sigma_i(t+\Delta t)= \sigma_i(t) \times \prod_j s_{ij}.
\end{equation}

\subsection{Defining a ruler}

As already discussed before \cite{martins15b}, defining opinion particles is an attempt to see how much we can recover from physical theories by starting from an opinion dynamics framework. And, if we assume $x_i$ is just a representation of an opinion, its actual value might have no physical meaning, except in comparison to the other averages and standard deviations in the system. More than that, from the point of view of a given particle $i$, it makes sense to describe its own coordinate system as centered where it is, so that, in the system $XI$ of particle $i$, we could have $XI_j(t)= x_j(t)- x_i(t)$.

We can take this one step further. For particle $i$, not only it makes sense to make itself the center of the coordinates, but it can use its own opinions to define its ruler, from where all sizes are measured. Indeed, particle $i$ has a natural definition of distances in terms of its uncertainty, measured by the standard deviation $\sigma_i$. That is, from the point of view of a particle $i$, the natural coordinate system $XI$ would actually be given by
\begin{equation}\label{eq:renorm}
	XI_j(t)= \frac{x_j(t)- x_i(t)}{\sigma_i(t)}.
\end{equation}

In the model here, we update both means and standard deviations for every particle using Equations \ref{eq:meancont} and \ref{eq:uncertcont}. From that, we can observe how the opinion particles evolve both in the original coordinates as well as in the natural coordinates defined by Equation \ref{eq:renorm}.

\section{Simulation results}

The synchronous model was implemented for $N=10$, $N=20$, and $N=50$ particles. In each implementation, all particles started, unless stated otherwise, at a random position in the $[0,1]$ interval. The results showed a substantial amount of variation, with the final position of each particle depending very strongly on where it started and the position of the other particles. To understand how the dynamics work, a central agent, labeled as $i=1$ was chosen to start at the middle of the range at $x_1 = 0.5$. A second agent, $i=2$ was chosen as probe, starting at a fixed distance $d$ of agent $i=1$, that is, $x_2 = 0.5 + d$.

\begin{figure}
	\centering
	\begin{tabular}{cc}
		\includegraphics[scale=.35]{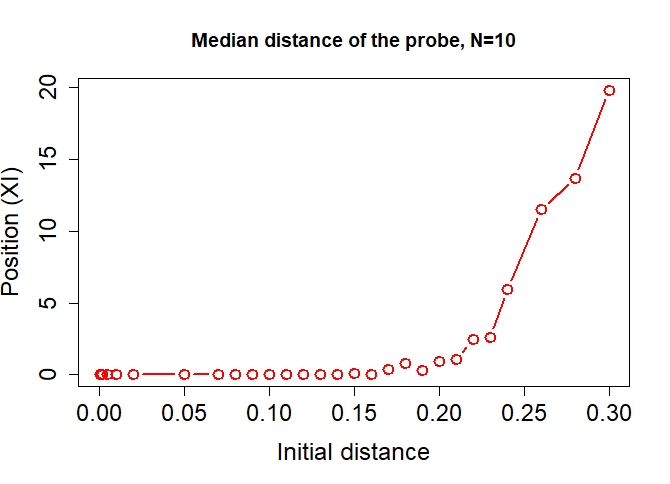}&
		\includegraphics[scale=.35]{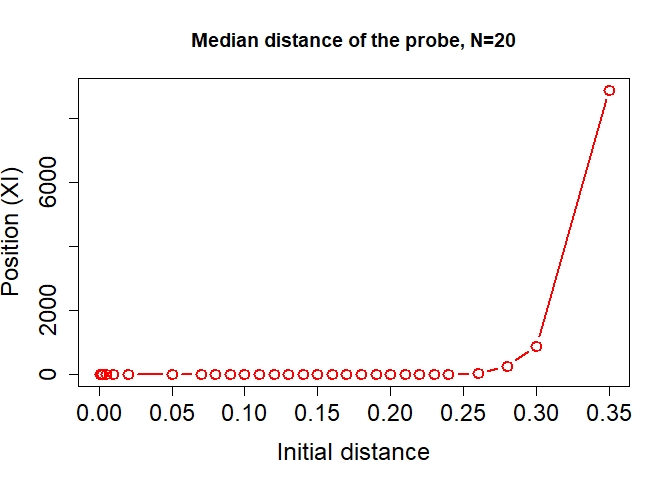}\\
		\includegraphics[scale=.35]{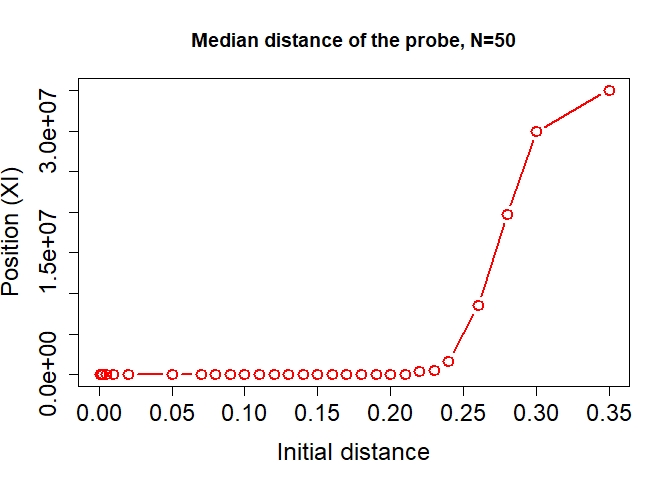}&
		\includegraphics[scale=.35]{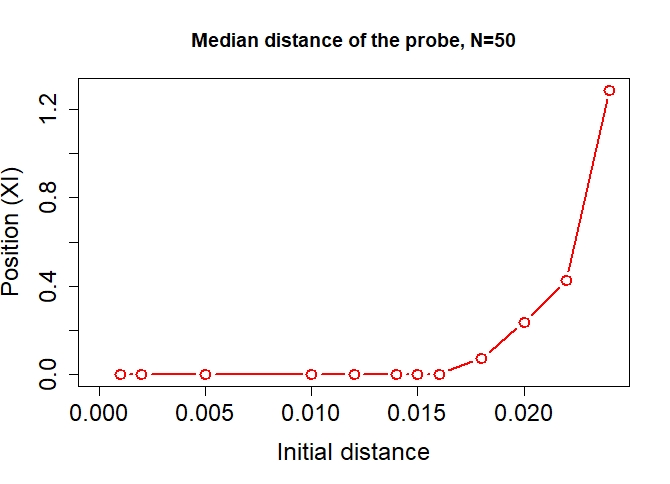}
	\end{tabular}
	\caption{Median final position of the probe particle after 50 time steps as a function of the initial distance between the probe and the particle used to define the natural coordinates. {\it Top line}: At left, $N=10$ and, at right, $N=20$. The result shown corresponds to the median of 10,000 realizations for each point.  {\it Bottom line}: Results for $N=50$ are shown in both panels. The right panel corresponds to the median of 10,000 realizations for each point while the right panel, detailing the transition between attraction and repulsion, corresponds to 50,000 realizations.
	}\label{fig:medians}
\end{figure}

As an example of the effects of the model, Figure~\ref{fig:medians} shows the median results for the final position of agent $i=2$ as a function of its starting distance to $i=1$, $d$. Medians were chosen instead of averages due to the large variability in the results. In each panel, the particles interacted synchronously with every other particle for 50 time-steps. Initial conditions were $\sigma_i =0.05$ and $p=0.005$ for all agents. The top two panels show the results for $N=10$ (left) and $N=20$ (right). In contrast, both bottom panels show the $N=50$ case. The one in the right shows in more detail the distances where the particle median position shows a transition between an attractive and a repulsive regime.

Such a transition can be observed for all the cases shown. What we see is that when particles start nearby, they tend to move towards each other. But when the probe particle begins further away, it will diverge, driven out by the decrease in the standard deviation $\sigma_i$. As the ruler scale becomes smaller, the distances it measures increase. And, as we see when there are more particles, such as in the $N=50$ case, that distancing can become quite dramatic. It is fascinating to notice that the more crowded the initial space is, the more distant the probe will end up from the central particle. That is, as long as it starts at a significant enough initial distance.

It is worth noticing that, without the natural coordinates $XI$, the effect of Equation~\ref{eq:averagedecept} is always attraction. As $p^*$ tends to zero for very large distances, the attraction can become small, but it never changes sign. The only cause for the apparent repulsive force is the redefinition of sizes caused by the update of $\sigma_1$.

\begin{figure}
	\centering
	\includegraphics[scale=.7]{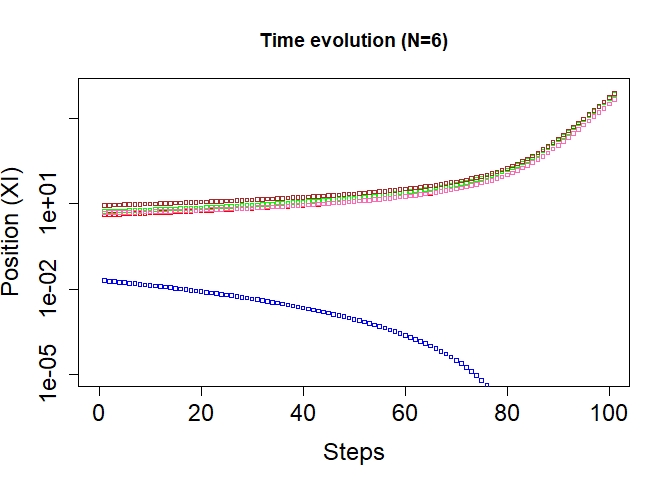}
	\caption{Time evolution of the other five particles in a system with $N=6$ particles, when seen in the natural coordinates of agent $i=1$.
	}\label{fig:opevolve}
\end{figure}

As another way to visualize how the dynamics of this model works, Figure~\ref{fig:opevolve} shows the time evolution of one single realization of a $N=6$ system. In the graphic, each set of points correspond to the evolution of the natural position $XI_j$ of all five non-central particles. The probe was chosen to be close to the central particle, with $d=0.001$, and its trajectory can be seen in blue. We can see how all other particles start moving away from the center and how they speed up as  $\sigma_1$ gets smaller. As a matter of fact, on some realizations, $\sigma_1$ can become too small. And it is also possible to observe cases where $XI_2$ becomes exactly zero due to limitations in the computing precision. When either of those cases happens, the implementation can generate non-sense, with the close particles jumping away for a while, due to truncation errors. Such cases must, of course, be disregarded.

	\section{Discussion}
	
	Here, we have developed further the idea \cite{martins15b} that models for opinion dynamics, especially those inspired by Bayesian methods \cite{Martins2021}, can be used as inspiration for models in physics. Extending a previous model for continuous opinions \cite{martins08c} to continuous time has lead to interesting interactions. Equation~\ref{eq:averagedecept} already showed, even out of the continuous-time limit,  a behavior, when $p=1$, that can be described as similar to asymptotic freedom \cite{Gross1973}.

	However, the most exciting characteristic of this model is to illustrate how an attractive interaction can cause distant objects to move away. Such a repulsive effect can be observed when the attraction interaction is also associated with measuring distances. That way, if the size of the ruler induced by the interaction decreases with time, close particles might still feel the interaction as attractive while distant ones are repelled. In that sense, the observed behavior can be thought of as a toy model for dark energy \cite{Peebles2003}. As there are no quantum or relativistic effects in this model, it should be seen as an initial exploration. On the other hand, as distances are actually defined in the model, it might be worth exploring in the future what kind of space-times we can obtain if we start from opinion particles and use their interactions as the base upon which geometry is defined.

	\section{Acknowledgments}

This work was supported by the Funda\c{c}\~ao de Amparo a Pesquisa do Estado de S\~ao Paulo (FAPESP) under grant  2019/26987-2.

%
\bibliographystyle{unsrt}
\bibliography{biblio}

\begin{thebibliography}{10}

\bibitem{castellanoetal07}
C.~Castellano, S.~Fortunato, and V.~Loreto.
\newblock Statistical physics of social dynamics.
\newblock {\em Reviews of Modern Physics}, 81:591--646, 2009.

\bibitem{latane81a}
B.~Latan\'e.
\newblock The psychology of social impact.
\newblock {\em Am. Psychol.}, 36:343--365, 1981.

\bibitem{galametal82}
S.~Galam, Y.~Gefen, and Y.~Shapir.
\newblock Sociophysics: A new approach of sociological collective behavior:
  Mean-behavior description of a strike.
\newblock {\em J. Math. Sociol.}, 9:1--13, 1982.

\bibitem{galammoscovici91}
S.~Galam and S.~Moscovici.
\newblock Towards a theory of collective phenomena: Consensus and attitude
  changes in groups.
\newblock {\em Eur. J. Soc. Psychol.}, 21:49--74, 1991.

\bibitem{sznajd00}
K.~Sznajd-Weron and J.~Sznajd.
\newblock Opinion evolution in a closed community.
\newblock {\em Int. J. Mod. Phys. C}, 11:1157, 2000.

\bibitem{deffuantetal00}
G.~Deffuant, D.~Neau, F.~Amblard, and G.~Weisbuch.
\newblock Mixing beliefs among interacting agents.
\newblock {\em Adv. Compl. Sys.}, 3:87--98, 2000.

\bibitem{martins08a}
Andr\'e C.~R. Martins.
\newblock Continuous opinions and discrete actions in opinion dynamics
  problems.
\newblock {\em Int. J. of Mod. Phys. C}, 19(4):617--624, 2008.

\bibitem{martins12b}
Andr\'e C.~R. Martins.
\newblock Bayesian updating as basis for opinion dynamics models.
\newblock {\em AIP Conf. Proc.}, 1490:212--221, 2012.

\bibitem{galam12a}
Serge Galam.
\newblock {\em Sociophysics: A Physicist's Modeling of Psycho-political
  Phenomena}.
\newblock Springer, 2012.

\bibitem{schweitzerholyst00}
F.~Schweitzer and J.A. Holyst.
\newblock Modelling collective opinion formation by means of active brownian
  particles.
\newblock {\em Eur. Phys. J. B}, 15:723--732, 2000.

\bibitem{Wang2017}
Peng Wang, Jia Song, Jie Huo, Rui Hao, and Xu-Ming Wang.
\newblock Towards understanding what contributes to forming an opinion.
\newblock {\em Int. J. Mod. Phys. C}, 28(11):1750135, November 2017.

\bibitem{Wang2021}
Peng Wang, Feng-Chun Pan, Jie Huo, and Xu-Ming Wang.
\newblock Non-equilibrium diffusion in a particle system and the correspondence
  to understanding the propagation of public opinion.
\newblock {\em Nonlinear Dynamics}, 2021.

\bibitem{martins15b}
Andr\'e C.~R. Martins.
\newblock Opinion particles: Classical physics and opinion dynamics.
\newblock {\em Physics Letters A}, 379(3):89--94, 2015.
\newblock arXiv:1307.3304.

\bibitem{galammartins15a}
Serge Galam and Andr\'e C.~R. Martins.
\newblock Two-dimensional ising transition through a technique from two-state
  opinion-dynamics models.
\newblock {\em Phys. Rev. E}, 91(1):012108, 2015.
\newblock arXiv:1303.2957.

\bibitem{Martins2021}
Andre C.~R. Martins.
\newblock Agent mental models and bayesian rules as a tool to create opinion
  dynamics models.
\newblock arXiv:2106.00199, 2021.

\bibitem{jaynes03}
E.T. Jaynes.
\newblock {\em Probability Theory: The Logic of Science}.
\newblock Cambridge, Cambridge University Press, 2003.

\bibitem{caticha04a}
A.~Caticha.
\newblock Relative entropy and inductive inference.
\newblock In G.~Erickson and Y.~Zhai, editors, {\em Bayesian Inference and
  Maximum Entropy Methods in Science and Engineering}, number~75 in AIP Conf.
  Proc. 707, 2004.
\newblock arXiv.org/abs/physics/0311093.

\bibitem{catichapreuss04a}
A.~Caticha and R.~Preuss.
\newblock Maximum entropy and bayesian data analysis: Entropic prior
  distributions.
\newblock {\em Phys. Rev. E}, 70:046127, 2004.

\bibitem{catichagiffin07a}
Ariel Caticha and Adom Giffin.
\newblock Updating probabilities.
\newblock In A.~Mohammad-Djafari, editor, {\em Bayesian Inference and Maximum
  Entropy Methods in Science and Engineering}, volume 872 of {\em AIP Conf.
  Proc.}, page~31, 2007.

\bibitem{goyal12a}
Philip Goyal.
\newblock Information physics: Towards a new conception of physical reality.
\newblock {\em Information}, 3:567--594, 2012.

\bibitem{Leandre2018}
Remi Léandre, Sean~Alan Ali, Carlo Cafaro, Steven Gassner, and Adom Giffin.
\newblock An information geometric perspective on the complexity of macroscopic
  predictions arising from incomplete information.
\newblock {\em Advances in Mathematical Physics}, 2018:2048521, 2018.

\bibitem{martins08c}
Andr\'e C.~R. Martins.
\newblock Bayesian updating rules in continuous opinion dynamics models.
\newblock {\em Journal of Statistical Mechanics: Theory and Experiment},
  2009(02):P02017, 2009.
\newblock arXiv:0807.4972v1.

\bibitem{Maciel2020}
Marcelo~V. Maciel and Andr\'e C.~R. Martins.
\newblock Ideologically motivated biases in a multiple issues opinion model.
\newblock {\em Physica A}, page 124293, 2020.
\newblock https://arxiv.org/abs/1908.10450.

\bibitem{hegselmannkrause02}
R.~Hegselmann and U.~Krause.
\newblock Opinion dynamics and bounded confidence models, analysis and
  simulation.
\newblock {\em Journal of Artificial Societies and Social Simulations}, 5(3):3,
  2002.

\bibitem{Peebles2003}
P.~J. {Peebles} and Bharat {Ratra}.
\newblock {The cosmological constant and dark energy}.
\newblock {\em Reviews of Modern Physics}, 75(2):559--606, April 2003.

\bibitem{VazMartins2010}
T.~Vaz~Martins, M.~Pineda, and R.~Toral.
\newblock Mass media and repulsive interactions in continuous-opinion dynamics.
\newblock {\em EPL (Europhysics Letters)}, 91(4):48003, 2010.

\bibitem{galam04}
S.~Galam.
\newblock Contrarian deterministic effect: the hung elections scenario.
\newblock {\em Physica A}, 333:453--460, 2004.

\bibitem{martinskuba09a}
Andr\'e C.~R. Martins and Cleber~D. Kuba.
\newblock The importance of disagreeing: Contrarians and extremism in the coda
  model.
\newblock {\em Adv. Compl. Sys.}, 13:621--634, 2010.

\bibitem{Khalil2019}
Nagi Khalil and Raúl Toral.
\newblock The noisy voter model under the influence of contrarians.
\newblock {\em Physica A: Statistical Mechanics and its Applications},
  515:81--92, 2019.

\bibitem{Gross1973}
David~J. Gross and Frank Wilczek.
\newblock Ultraviolet behavior of non-abelian gauge theories.
\newblock {\em Phys. Rev. Lett.}, 30:1343--1346, Jun 1973.

\end{thebibliography}

\end{document}